# Electrical Control of Surface Acoustic Waves

Linbo Shao[1,2,*,†], Di Zhu[1,3,†], Marco Colangelo[4], Daehun Lee[5], Neil Sinclair[1,6], Yaowen Hu[1,7], Peter T. Rakich[8], Keji Lai[5], Karl K. Berggren[4], Marko Lončar[1,*]

[1]John A. Paulson School of Engineering and Applied Sciences, Harvard University, Cambridge, Massachusetts 02138, USA
[2]Bradley Department of Electrical and Computer Engineering, Virginia Tech, 1185 Perry Street, Blacksburg, Virginia 24061, USA
[3]Institute of Materials Research and Engineering, Agency for Science, Technology and Research (A*STAR), Singapore 138634, Singapore
[4]Research Laboratory of Electronics, Massachusetts Institute of Technology, Cambridge, Massachusetts 01239, USA
[5]Department of Physics, University of Texas at Austin, Austin, Texas 78712, USA
[6]Division of Physics, Mathematics and Astronomy, and Alliance for Quantum Technologies (AQT), California Institute of Technology, Pasadena, California 91125, USA
[7]Department of Physics, Harvard University, Cambridge, Massachusetts 02138, USA
[8]Department of Applied Physics, Yale University, New Haven, Connecticut 06520, USA
*Correspondence to: shaolb@vt.edu (L.S.); loncar@seas.harvard.edu (M.L.)
†These authors contributed equally to this work.

## Abstract

Acoustic waves at microwave frequencies have been widely used in wireless communication and recently emerged as versatile information carriers in quantum applications. However, most acoustic devices are passive components, and dynamic control of acoustic waves in a low-loss and scalable manner remains an outstanding challenge, which hinders the development of phononic integrated circuits. Here we demonstrate electrical control of traveling acoustic waves on an integrated lithium niobate platform at both room and millikelvin temperatures. We modulate the phase and amplitude of the acoustic waves and demonstrate an acoustic frequency shifter by serrodyne phase modulation. Furthermore, we show reconfigurable nonreciprocal modulation by tailoring the phase matching between acoustic and quasi-traveling electric fields. Our scalable electro-acoustic platform comprises the fundamental elements for arbitrary acoustic signal processing and manipulation of phononic quantum information.

## Main Text

Acoustic waves in solids are the basis for numerous applications[1] including microwave filters, oscillators, delay lines, and sensors. They are also emerging as versatile interfaces between quantum systems such as superconducting circuits[2-8], defect centers[9,10], and optical devices[11-15]. Compared to gigahertz electromagnetic waves, acoustic waves feature five-orders-of-magnitude shorter wavelength and do not radiate into free-space. This therefore allows coherent information processing and manipulation in an ultra-compact footprint with negligible crosstalk between devices and with the environment. For these reasons, on-chip phononic systems have emerged as a promising candidate for quantum computing and storage[15-17]. A phononic integrated circuit requires a few essential functionalities, including low-loss waveguiding of acoustic waves, efficient transduction from and to electromagnetic waves at microwave frequencies, and, importantly, dynamic routing and modulating of traveling acoustic waves. Integrated acoustic waveguides have been realized using suspended structures[18,19], two-dimensional phononic crystals[20], and high acoustic velocity substrates[21,22]. Efficient transduction between acoustic waves and electromagnetic waves have also been demonstrated, leveraging piezoelectric coupling with microwaves[18,19,21,22] and optomechanical coupling with light[20]. However, active control on phase, amplitude, frequency, and nonreciprocity of acoustic waves in a low-loss and scalable manner remains an outstanding challenge, which hinders the development of acoustic integrated circuits, especially at millikelvin temperatures.

Among previous exploration to control on-chip acoustic waves, approaches based on acoustic four-wave mixing[21] and nonlinear mechanical cavities[19] were inefficient and required large acoustic powers due to the weak nonlinearity in the elastic response of most materials. Alternatively, directional amplification and manipulation of sub-GHz acoustic waves have been demonstrated using electrical amplifiers[23] and semiconductor acoustoelectric effects[24-26]. However, their reliance on thermal excitations of charge carriers in semiconductors inevitably induces loss and noise of the acoustic wave, and is incompatible with cryogenic temperatures, which precludes their use in quantum applications. On the other hand, strategies used to achieve acoustic nonreciprocity based on nonlinear materials[27], circulating fluids[28], water-submerged phononic crystals[29], deformed water-air interfaces[30], and optomechanics[31] were all limited to acoustic frequencies below a few megahertz, which renders them unsuitable for applications that require microwave acoustic frequencies. Furthermore, approaches using ferromagnetic materials[32] require a magnetic field and therefore are inconvenient for co-integration with superconducting circuits or spins in solids.

Here we demonstrate electrical control of the fundamental degrees of freedom of acoustic waves on an integrated lithium niobate (LN) platform[33]. Modulation of the acoustic waves is enabled by the electro-acoustic effect (also known as third-order piezoelectric effect)[34]. The electro-acoustic effect is a parametric process that occurs at both room and millikelvin temperatures and is analogous to electro-optic effect widely utilized to control the phase and amplitude of optical signals. It describes the change in the elasticity of a solid due to an applied electric field, which results in a change of the phase velocity of traveling acoustic waves. The electro-acoustic effect is characterized by the third-order piezoelectric tensor $d$. The change of elastic constants $\Delta c$ due to the applied electric field $E$ is given by $\Delta c_{ij} = d_{kij} E_k$, where $i, j, k$ can take values of 1 to 3, corresponding to crystal $X$, $Y$ and $Z$ directions, and $d_{kij}$ is subject to the material symmetry[35]. However, since this electro-acoustic effect is relatively weak, bulk components show small phase changes[34] and are unsuitable for practical applications. Here we overcome this limitation by confining the acoustic wave to a wavelength-scale acoustic waveguide and placing the modulation electrodes closely across the waveguide. Doing so drastically enhances electrical-acoustic interaction and, therefore, allows full π-phase shift to be achieved.

**Electro-acoustic Modulator**
Our electro-acoustic modulators are fabricated on an X-cut LN substrate (Fig. 1a). The acoustic waveguide is formed by creating a 10 μm slot inside thin silicon nitride (SiN) film deposited on top of LN (Figs. 1b and 1c). Since the acoustic velocity (index) of SiN is greater (smaller) than that of LN, a Rayleigh-type acoustic mode is confined (Fig. 1b). Inter-digital transducers (IDT) are used to electrically excite and detect microwave acoustic waves. The pitch of the IDT finger electrodes is 650 nm and equal to the half wavelength of the acoustic waves at 2.5 GHz. To optimize the transduction efficiency, the width of the IDT (75 μm) is designed to be larger than the acoustic waveguide (10 μm), and tapered waveguide structures are used to couple the wave into the acoustic waveguide. Importantly, the waveguide is oriented along 30° angle with respect to the crystal Z axis, as this direction features the smallest acoustic velocity on the X-cut surface and thus provides the best acoustic wave confinement (Extended Data Fig. 1). Finally, aluminum electrodes are deposited on the SiN layer and used to apply the electric field needed for acoustic index modulation. Low loss is critical for realizing large-scale phononic integrated circuits. We measure the propagation loss of the acoustic waveguide at different temperatures down to 1 K (Fig. 1d). At room temperature (300 K) and under vacuum, the acoustic propagation loss is $\alpha = 17$ dB/cm. It decreases with lowered temperature to $\alpha = 5$ dB/cm at liquid nitrogen temperature (77 K) and <1 dB/cm at 1.3 K. These values are consistent with those measured using acoustic cavities on LN[21,36]. Thermoelastic dissipation[37] is the likely source of loss at temperatures above 1 K.

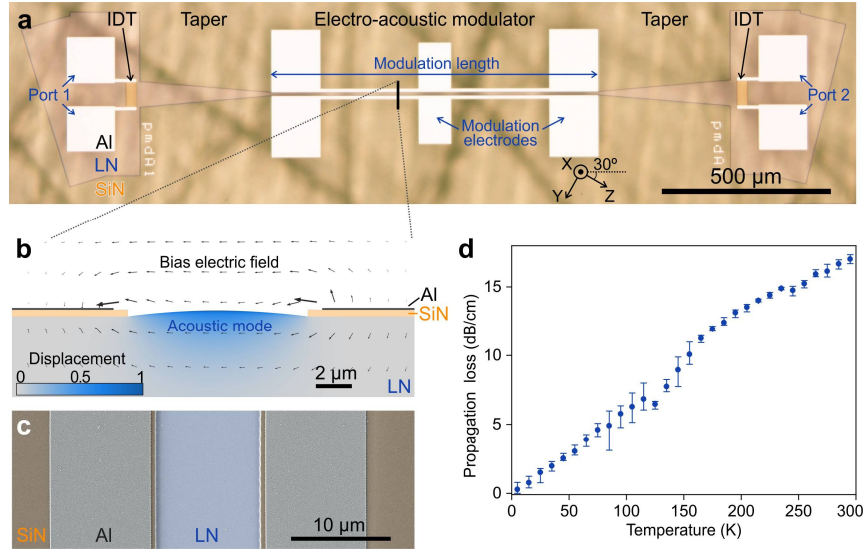

**Fig. 1 | Lithium niobate (LN) electro-acoustic platform. a**, Optical micrograph of a fabricated device. Bright regions are aluminum (Al). The etched silicon nitride (SiN) layer showing dark boundaries is used to define the acoustic waveguide and regions where the interdigital transducers (IDT) are fabricated. IDTs are used to excite and detect the acoustic waves. The device is on an X-cut LN substrate, and the acoustic waveguide is at 30° angle with respect to the crystal Z-axis (coordinates indicated). The out-of-focus dark lines are scratches on the back side of the chip, which do not affect the device. The modulation length of the device shown in **a** is 1 mm, while devices with different modulation lengths are used throughout this work. **b**, Cross-section of the acoustic waveguide in the modulation region. The normalized displacement field intensity (blue shading) shows the simulated fundamental acoustic mode. The displacement field is exaggerated for visualization. Arrows indicate the simulated electric field direction and magnitude due to a bias voltage on the modulation electrodes. **c**, False-colored scanning electron microscope image of the acoustic waveguide. **d**, Measured propagation loss of the acoustic waveguide as a function of temperature. The error bar indicates the maximum and minimum losses measured.

## Phase Modulation of Acoustic Waves

We measure the performance of a 1-cm-long electro-acoustic phase modulator by inputting an acoustic wave at carrier frequency $f_c \sim 2.5$ GHz and detecting phase and amplitude of the modulated acoustic wave (Fig. 2a). The insertion loss of the device, measured from a microwave signal applied to one IDT and detected after the other, is 10 dB at cryogenic temperature. This loss is dominated by the tapers that guide acoustic waves to the waveguide and the symmetric IDTs that excite and collect acoustic waves bidirectionally, which result in a 3 dB loss at each IDT. The insertion loss could be further reduced by employing unidirectional IDTs[38]. At room temperature, the insertion loss increases by 25 dB due to higher propagation loss and a lower efficiency of IDTs (Extended Data Fig. 2). We used transmission-mode microwave impedance microscopy (TMIM)[39,40], a scanning probe technique that coherently measures the profiles of traveling acoustic waves near the output of the modulator waveguide. We observe a ~π/2 phase shift of the acoustic wave when a DC bias voltage is applied on the modulation electrode (Fig. 2b). This measurement of the traveling-wave profile confirms that the acoustic wave is being directly modulated.

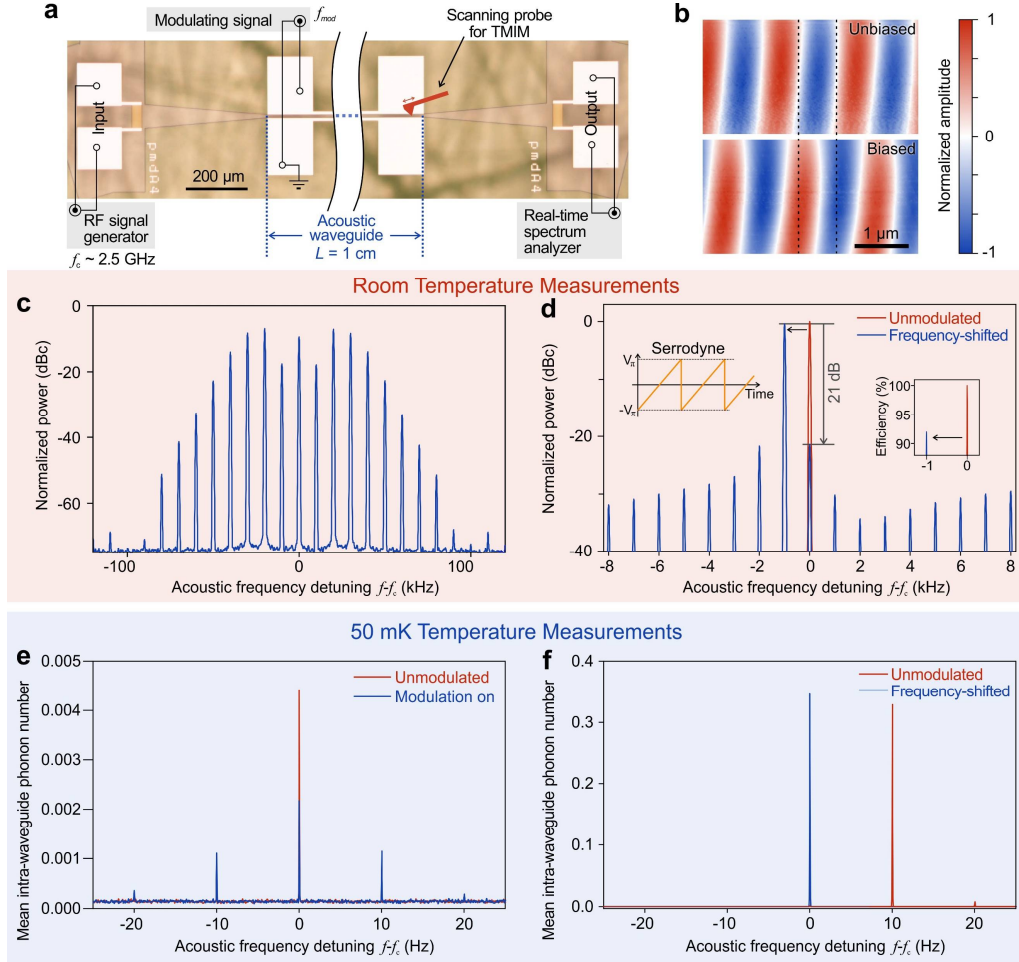

**Fig. 2 | Electro-acoustic phase modulation**. **a**, Experimental setup for characterizing the electro-acoustic phase modulator at room temperature. A signal generator is used to excite the acoustic wave via an IDT at carrier frequency $f_c$. A modulating signal with frequency $f_{mod}$ is applied to the modulation electrodes, and a real-time spectrum analyzer is used to detect the phase and amplitude of transmitted acoustic signal detected by an identical IDT. In the transmission-mode microwave impedance microscopy (TMIM), a scanning probe is used to detect the acoustic field. **b**, TMIM images show phase shift of the acoustic wave due to an applied bias voltage on the modulation electrodes. The scanning region is located at the center of the waveguide near the output of the modulator. **c**, Modulation of an acoustic wave by a 10 kHz sine wave with $V_{pp} = 2.3\ V_\pi$. **d**, Serrodyne frequency shift of an acoustic wave by 1 kHz. This is achieved by applying a repeating linear voltage ramp (serrodyne) with $V_{pp} = 2\ V_\pi$ (Left inset). Right inset plots the acoustic powers in a linear scale, showing an efficiency of 92%. The spectral powers in **c** and **d** are normalized to the unmodulated signal received by the spectrum analyzer. Measurements in **b**, **c** and **d** are conducted at room temperature and using $f_c$ = 2.483 GHz. **e**, Modulation of an acoustic wave by a 10 Hz sine wave with $V_{pp}$ = 140 V. Unmodulated signal is also shown. The input microwave signal is attenuated such that the mean phonon number of the acoustic wave in the waveguide is much less than 1. **f**, Serrodyne frequency shift of 10 Hz at the single-phonon level with an efficiency of 94.8%. This is achieved by applying a repeating linear voltage ramp (serrodyne) with $V_{pp} = 2\ V_\pi$. **e** and **f** are measured at 50 mK temperature with $f_c$ = 2.528 GHz. All results presented in this figure are measured from the same device with a modulation length of 1 cm.

To reliably measure the half-wave voltage $V_\pi$ of the modulator, we apply a sinusoidal signal at $f_{\text{mod}} = 10$ kHz on the phase modulator electrode and analyze the output by a real-time spectrum analyzer. The phase change increases linearly with the amplitude of modulating signal (Extended Data Fig. 3). When the peak-to-peak voltage ($V_{\text{pp}}$) of the modulating signal reaches 53 V, a π-phase change of the received acoustic wave is observed at room temperature, inferring a $V_\pi = 53$ V. At 1.3 K, the peak transmission frequency shifts to 2.532 GHz, and the $V_\pi$ increases to 135 V (Extended Data Fig. 3) as the material properties vary with the temperature. Despite the increased $V_\pi$ at the cryogenic temperature, an important figure of merit for the modulator, the product between its half-wave voltage $V_\pi$, length $L$, and propagation loss α, i.e., $V_\pi L\, α$, is reduced by a factor of 7 (from 900 V·dB to 120 V·dB) at 1.3 K owing to reduced propagation loss compared to room temperature. The $V_\pi$ of the phase modulator could be further reduced by a factor of 2 by using narrower acoustic waveguides, which is possible using materials with higher acoustic contrast. In addition, we measure the 3-dB bandwidth to be 110 kHz and observe zero modulation at $f_{\text{mod}} = 336$ kHz when the period of the modulating signal matches the traveling time of the acoustic wave (Extended Data Fig. 4).

Next, we demonstrate two proof-of-concept applications – electro-acoustic frequency comb and acoustic frequency shifting. By driving the phase modulator with a 10 kHz sinusoidal signal of $V_{\text{pp}}=2.3\ V_\pi$, we generate 19 equidistant frequency comb lines centered at $f_c = 2.483$ GHz (Fig. 2c). The frequency comb coherently generates new frequencies and could be useful for short acoustic pulse generations and frequency domain information processing. Additionally, the ability to modulate over a full 2π phase allows us to demonstrate acoustic frequency shifting using a serrodyne approach. Specifically, by applying a repeating linear voltage ramp signal at the frequency of 1 kHz and $V_{\text{pp}}$ of 2 $V_\pi$, the modulated acoustic wave experiences an approximately linear phase ramp in time, which results in change (shift) in frequency of the acoustic wave. We measure on-chip frequency shift efficiency of 92% (Fig. 2d), defined as the ratio of detected acoustic power at the shifted frequency and the power of unmodulated acoustic wave. The carrier suppression of our frequency shifter is 21 dB.

To assess our electro-acoustic modulator for quantum applications, we demonstrate coherent modulation of single-phonon-level acoustic waves at 50 mK (see Methods and Extended Data Fig. 5). We attenuate the input microwave signal so that the mean phonon number of the field in the acoustic waveguide is much less than one. Basic functionalities of the electro-acoustic modulator are preserved at low temperature: symmetric sidebands are observed when applying a 10 Hz sinusoidal modulation signal (Fig. 2e) and serrodyne frequency shifting, with an efficiency of 94.8%, is realized by applying a repeating linear voltage ramp at 10 Hz using $V_{\text{pp}}$ of 2 $V_\pi$ (Fig. 2f). Importantly, we experimentally verify that the modulation process adds less than one noise phonon (Extended Data Fig. 6).

**Amplitude Modulation of Acoustic Waves**
We achieve amplitude modulation of acoustic wave by constructing an acoustic Mach-Zehnder interferometer (MZI) in the push-pull configuration (Fig. 3a). The input acoustic wave is split equally between two MZI arms. As the electric fields applied over the waveguides are in opposite directions, the induced elasticity changes are also with opposite signs, and thus the two split waves experience opposite phase shifts as they propagate in each arm. The phase difference is controlled by the voltage applied to the electrodes. When the two waves recombine, the acoustic interference yields an amplitude modulation. Maximum output amplitude occurs when the phase difference between the two paths is zero (no voltage applied) or an even integer number of π, while the minimum amplitude occurs when the phase difference is an odd integer number of π. We measure $V_\pi$ = 29 V at quasi-DC frequency (100 Hz) for an 8-mm-long acoustic Mach-Zehnder modulator. The extinction ratio between the maximum and minimum output power is over 15 dB (Fig. 3b), likely limited by the fabrication imperfection of the Y-splitter or the existence of higher-order modes in the waveguide. When the modulator is biased at the quadrature point (i.e., 50% transmission), the amplitude of the output acoustic waves follows the small input signal accordingly (Fig. 3c).

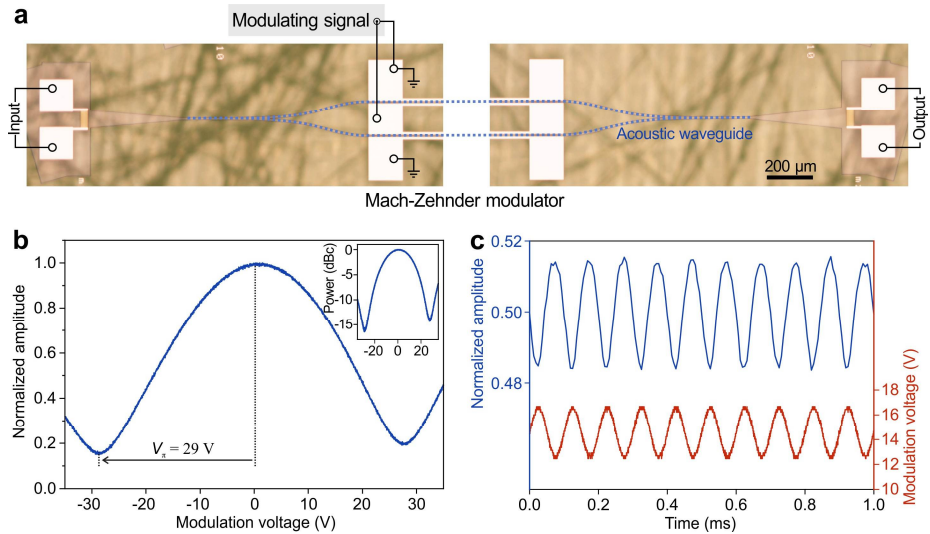

**Fig. 3 | Electro-acoustic amplitude modulation. a**, Schematic of the electro-acoustic Mach-Zehnder modulator. The modulating signal is applied at the middle electrode while two outer electrodes are ground, and thus two acoustic waveguides experience electric fields with opposite polarities. **b**, Measured output acoustic amplitude (main figure) and power (Inset) with slowly varying modulating voltage to determine the $V_\pi$ of the modulator. **c**, Measured acoustic amplitude with a weak modulating signal at $f_{mod}$ = 10 kHz and biased at 0.5 $V_\pi$. Measurements are at room temperature.

**Nonreciprocal Modulation of Acoustic Waves**

To achieve nonreciprocal transmission of acoustic waves, we employ a quasi-traveling electric field to break the symmetry of counter-propagating acoustic waves. By separating modulation electrode into three segments, we can control the wave number (momentum) of the quasi-traveling electric field by adjusting the relative phase of modulating signals applied to each electrode (Fig. 4a). This approach enables nonreciprocal acoustic phase modulation when the quasi-traveling modulating signal is phase matched with the traveling acoustic wave in one direction but mismatched in the opposite direction (Fig. 4b). Maximum nonreciprocity occurs when the signals applied to each succeeding modulation electrode segment are phase delayed by 120° and when the modulation frequency matches the total traveling time of the acoustic wave, *i.e.*, $1/f_{mod} = 3t_0$, where $t_0$ is the time for the acoustic wave to travel through one electrode segment. In this case, the forward propagating acoustic wave always experiences the same phase modulation when it traverses the electrodes and thus results in maximum modulation. On the other hand, the backward propagating wave effectively experiences a full (2π) modulation cycle, which results in no net phase change. To implement this concept, we fabricate such a nonreciprocal acoustic modulator with an overall electrode length of 1 cm and apply the required modulation frequency $f_{mod} = 1/(3t_0)$ = 336 kHz for maximum nonreciprocity. In this condition, we observe the presence (absence) of acoustic phase modulations in the forward (backward) propagation direction (Fig. 4c).

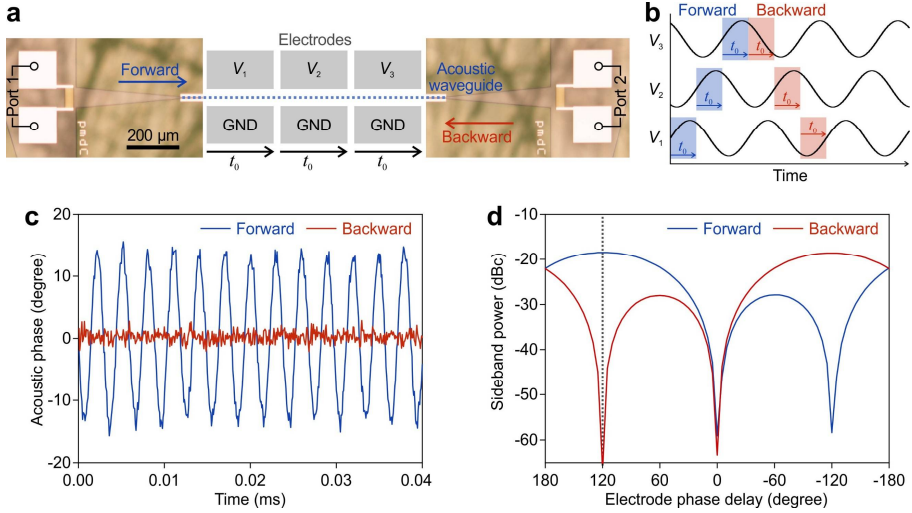

**Fig. 4. Nonreciprocal phase modulation of acoustic waves. a**, Schematic of the device used for nonreciprocal phase modulation. The modulation electrode is separated into three segments with independently controlled voltage $V_1$ to $V_3$. The acoustic traveling time through each segment is $t_0$. **b**, Illustration of nonreciprocal modulation of forward and backward propagating acoustic waves. The voltage applied on each segment of the electrodes is progressively delayed by 120°. For forward traveling acoustic wave, the phase accumulated in each segment is the same and results in net modulation. For backward traveling acoustic wave, the accumulated phase in each segment is different and can be designed to result in a net zero phase shift. **c**, Measured acoustic phases acquired for both forward and backward propagating acoustic waves. **d**, Measured modulation sideband power of the forward and backward propagating acoustic waves for varying phase delays between the voltages applied to the electrodes. The dashed line indicates the operating condition in **c** showing a nonreciprocity of >40 dB. The modulation frequency is $f_{\mathrm{mod}} = 1/(3t_0) = 336$ kHz in **c** and **d**. Measurements are at room temperature.

The acoustic nonreciprocity can be adjusted by the relative phase of the signals applied to the modulation electrodes (Fig. 4d). We observe a maximum nonreciprocity of over 40 dB in the modulation sideband power. Furthermore, we sweep the modulation frequency and relative phase delay of the applied voltages on the three electrodes. We observe maximum modulation when the traveling acoustic wave and the modulating signal are phase-matched and zero modulation when they are phase-mismatched by any positive integer number of $2\pi$ phases (Extended Data Fig. 7). With additional filters and couplers, acoustic isolators and circulators could be implemented based on our frequency shifter and nonreciprocal phase modulator.

**Conclusion and Discussion**

We demonstrate an integrated electro-acoustic platform that provides arbitrary control of on-chip traveling acoustic waves, namely their phase and amplitude. Compared with previous approaches, our electro-acoustic modulators show significant advantages in low-temperature compatibility, modulation efficiency, simplicity in fabrication, and scalability. Taken together, these advantages may enable large-scale integrated acoustic information processing systems. Furthermore, using advanced nanofabrication with tens of nanometers resolution, we could push the operation frequency of our device to tens of GHz, covering the 5G millimeter wave bands. We expect the half-wave voltage $V_\pi$ to decreases quadratically with higher frequency, as both the width of the acoustic waveguide and the acoustic wave number scale linearly with frequency. Our devices may find applications in emerging acoustically mediated quantum networks that allow connecting different solid-state systems and thus enable hybrid quantum networks that leverage the distinct functionalities of each system for quantum computing, communication, and sensing[41]. In particular, phase modulation is necessary for the control of coherent interactions and entanglement between solid state systems, dynamical routing and synchronization for addressability and error mitigation, compensating environmental changes such as local temperature drifts, or for compensating unavoidable detuning between quantum systems.

## Methods

**Design of the Electro-acoustic Modulators**

The interdigital transducers (IDTs) are optimized for maximum transduction between acoustic and electrical waves at 2.5 GHz. The aperture of the IDT is 75 μm, the pitch of the finger electrode is 650 nm, and the number of finger electrode pairs per IDT is 25. The maximum transmission between two IDTs at 2.5 GHz is -8 dB at room temperature, with -6 dB resulting from the symmetric design of the IDTs (-3 dB per IDT).

The adiabatic taper that connects an IDT to an acoustic waveguide is 400 μm long. The insertion loss per acoustic taper is about 5 dB at room temperature, as extracted from the measured transmission of devices with and without the tapered structure.

X-cut lithium niobate (LN) substrates are used for all devices. The direction of the acoustic waveguide is at an angle of 30° with respect to the crystal Z-axis (Fig. 1a). This direction features the slowest surface acoustic wave phase velocity using X-cut LN and thus leads to a well-confined acoustic mode for the waveguide (Extended Data Fig. 1). The modulating electric field applied across the waveguide is mainly in the crystal Y-direction. As most strain field of the guided acoustic mode is in the XZ component (corresponding to index 5 in the Voigt notation), our device employs a non-zero electro-acoustic modulation coefficient $d_{255} = d_{145}$.

**Device Fabrication**

A 400 nm-thick silicon nitride (SiN) layer is deposited using plasma-enhanced chemical vapor deposition on the X-cut LN substrate. The SiN layer is patterned using a direct-write lithography tool (Heidelberg Instruments MLA150) and etched using reactive-ion etching with carbon tetrafluoride ($CF_4$), sulfur hexafluoride ($SF_6$), and hydrogen ($H_2$) gases. The metal layer is patterned using electron-beam lithography (Elionix ELS-F125) with polymethyl methacrylate (PMMA) resist. A 115 nm-thick aluminum is deposited using electron-beam evaporation, followed by lift-off in 1-methyl-2-pyrrolidone (NMP) for more than 3 hours at 80 °C.

**Device Measurements**

The devices are mounted and wire-bonded to a printed circuit board (PCB). The transmission spectra of the devices are measured using a vector network analyzer (Keysight N5224A). For the modulation experiments, a microwave signal generator drives one IDT using a single-frequency tone around 2.5 GHz, and a real-time spectrum analyzer (RSA, Tektronix RSA 3303B) is connected to the other IDT. The RSA not only measures the power spectrum of the acoustic wave received by the IDT, but also demodulates the signal to provide real-time in-phase and quadrature data, which are converted to the phase and amplitude of the received signal. The microwave signal generator and the RSA are synchronized by a 10 MHz clock. An arbitrary waveform generator is used to provide any low-frequency modulating signals, and a 20 kHz-bandwidth voltage amplifier (Falco Systems, WMA-005) is used to provide a 20 times amplification in voltage when necessary. For nonreciprocal phase modulation, a four-channel arbitrary waveform generator (Tabor WS8104A-DST) is used to generate the three synchronized modulating signals with various phase delays.

**Low-temperature Measurements**

Two low-temperature setups are used in this work. For data in Fig. 1, Extended Data Figs. 2 and 3. We use a closed-cycle cyrostat (ICE Oxford) that reaches a base temperature of ~0.8 K. The temperature, transmission ($S_{21}$), and half-wave voltage of the devices are monitored continuously as the cryostat cools from room temperature. These measurements are repeated as the cryostat warms up. Cable losses are independently calibrated during a separate cooldown. To characterize the propagation loss of the SAWs, the transmission of two acoustic waveguides with different lengths is measured and compared. These waveguides are fabricated on the same chip, packaged on the same PCB, and connected using identical cables.

For data shown in Figs. 2e, 2f, and Extended Data Fig. 6, a 50 mK measurement setup is used (Extended Data Fig. 5). Our device is mounted on the mixing chamber plate of a dilution fridge (Bluefors) with a base temperature below 50 mK. The microwave input signal to the IDT of the electro-acoustic modulator is attenuated such that the mean phonon number on the 1-cm-long acoustic waveguide is less than one. The readout line includes a circulator, a high-electron-mobility transistor (HEMT) amplifier at 4 K, and two room-temperature low-noise amplifiers. The output signal is recorded using a real-time spectrum analyzer (RSA). The detection bandwidth of the RSA is set to 78 mHz to realize a high signal-to-noise ratio, but this detection bandwidth limits the span to 50 Hz. To detect the spectrum in this configuration of RSA, the modulation signal is set to 10 Hz. The modulation signal is supplied by a function generator followed by a high-voltage amplifier (Trek 2210). A controlled and isolated thermal source, consisting of a heater, temperature sensor, and a 30 dB attenuator, is installed on the input line before the electro-acoustic modulator to calibrate the readout gain and added noise[42]. We vary the temperature of this thermal source from 100 mK to 6.7 K and measure the output noise spectrum density. By comparing the measured noise spectrum density on the RSA against the calibrated thermal noise source, we extract a total readout gain of 91.49 dB, consistent with the specifications of the amplifiers and cables used.

We investigate the noise performance of our electro-acoustic modulator at 50 mK (Extended Data Fig. 6) by measuring the noise floor near the carrier frequency in three situations: (A) no signal is applied to the electro-acoustic modulator, (B) only the carrier microwave signal is applied, and (C) both carrier microwave signal and modulation signal are applied. When no signal is applied (Situation A), we measure a noise phonon number of $N_{tot,A} = 64.35 \pm 0.39$ quanta/s/Hz, which is mainly from the HEMT amplifier. When applying the carrier microwave signal (Situation B), we measure $N_{tot,B} = 64.23 \pm 0.36$ quanta/s/Hz. We further apply the modulation signal (Situation C) and observe $N_{tot,C} = 64.32 \pm 0.37$ quanta/s/Hz. As no observable difference (within error) in noise are measured, we conclude that our electro-acoustic modulation adds negligible noise and is thus suitable for quantum applications. This result agrees with the fact that our electro-acoustic modulation is a parametric process.

**Transmission-Mode Microwave Impedance Microscopy**
The acoustic-wave profiles in the main text are directly imaged using the transmission-mode microwave impedance microscopy (TMIM)[39,40], which is implemented on a commercial atomic-force microscopy (AFM) platform ParkAFM XE-70. The IDT is driven by a continuous microwave input signal (Anritsu MG 3692A), which launches the propagating surface acoustic wave. During the AFM scanning, the customized cantilever probe from PrimeNano Inc. picks up the GHz piezoelectric potential accompanying the acoustic wave. By using the same excitation frequency as the reference, the TMIM electronics demodulate the tip signal into a time-independent spatial pattern that is shown in Fig. 2B. Note that the TMIM image contains information on the phase of the propagating wave [39]. As a result, a lateral shift of the wave pattern indicates that the acoustic wave is modulated by the DC bias electric field. Due to the charging effects at interfaces between layers, a higher DC bias voltage is required to achieve the same phase shift as that of a modulating signal at $f_{mod}$=10 kHz.

**Measurement of the Modulation Bandwidth**
First, we measure the modulation bandwidth of the 1-cm-long phase modulator (Fig. 2). We apply a weak signal modulation ($V_{pp}$ = 0.2 $V_\pi$) with varied frequencies and measure the modulation efficiency. The modulation efficiency is indicated by the first sideband power generated by the phase modulation. Due to the phase mismatch between the slowly propagating acoustic wave and the fast-varying electrical modulating signals, we measure the 3-dB bandwidth to be 110 kHz for the 1-cm phase modulator and observe periodic variations of the sideband power as a function of modulation frequency (Extended Data Fig. 4). The modulation efficiency approaches zero every time when the modulation frequency is an integer multiple (N) of $f_{mod} = 336$ kHz. At these frequencies, the electric field oscillates exactly $N$ full cycles as the acoustic wave travels through the modulator, resulting in a vanishing cumulative modulation effect. The zero-modulation frequency is related to the modulation length $L$ and the acoustic group velocity $v_g$ by $f_{mod} = \frac{v_g}{L}$. Using this relationship, we infer an acoustic group velocity of $v_g = 3.36$ km/s, consistent with the simulated velocity of 3.38 km/s (Extended Data Fig. 1).

**Lithium Niobate Electro-acoustic Effect**
Hooke's law says that the force on a spring is proportional to its displacement. The square of the resonant frequency of a mass-spring system is equal to the ratio of the spring proportionality constant to the mass. Thus, tuning the spring constant also varies the resonance frequency of the spring.

Weakly-excited acoustic waves in solids follow a generalized Hooke's law that relates stress $\sigma$ and strain $\epsilon$ by an elasticity (stiffness) matrix $C$, which is a 6-by-6 matrix in Voigt notation. LN is of point group *3m*, which has a three-fold rotation symmetry about its Z axis and mirror symmetry on its X axis. With vanishing components of $C$ due to the symmetry of LN, the relation is

$$\begin{bmatrix}\sigma_1\\\sigma_2\\\sigma_3\\\sigma_4\\\sigma_5\\\sigma_6\end{bmatrix} = \begin{bmatrix}c_{11} & c_{12} & c_{13} & c_{14} & 0 & 0\\c_{12} & c_{11} & c_{13} & -c_{14} & 0 & 0\\c_{13} & c_{13} & c_{33} & 0 & 0 & 0\\c_{14} & -c_{14} & 0 & c_{44} & 0 & 0\\0 & 0 & 0 & 0 & c_{44} & c_{14}\\0 & 0 & 0 & 0 & c_{14} & \frac{1}{2}(c_{11}-c_{12})\end{bmatrix}\begin{bmatrix}\epsilon_1\\\epsilon_2\\\epsilon_3\\\epsilon_4\\\epsilon_5\\\epsilon_6\end{bmatrix}.$$

The elasticity matrix of LN can be varied by an applied electric field, $\Delta c_{ij} = d_{kij} E_k$, where $i,j$=1..6, $k$=1, 2, 3, $\mathbf{E}$ is the applied electric field, and $\mathbf{D}$ ($d_{kij}$) is the third-order piezoelectric tensor[34]. Subject to the symmetry of LN, $\mathbf{D}$ has the following form,

$$d_{1ij} = \begin{bmatrix} 0 & 0 & 0 & 0 & d_{115} & d_{116} \\ 0 & 0 & 0 & 0 & d_{125} & d_{126} \\ 0 & 0 & 0 & 0 & d_{135} & d_{136} \\ 0 & 0 & 0 & 0 & d_{145} & \frac{1}{2}(d_{115} - d_{125}) \\ d_{115} & d_{125} & d_{135} & d_{145} & 0 & 0 \\ d_{116} & d_{126} & d_{136} & \frac{1}{2}(d_{115} - d_{125}) & 0 & 0 \end{bmatrix}$$

$$d_{2ij} = \begin{bmatrix} \frac{1}{2}(d_{116} + 3\,d_{126}) & \frac{1}{2}(d_{116} - d_{126}) & d_{136} & d_{125} & 0 & 0 \\ \frac{1}{2}(d_{116} - d_{126}) & -\frac{1}{2}(3d_{116} + d_{126}) & -d_{136} & d_{115} & 0 & 0 \\ d_{136} & -d_{136} & 0 & d_{135} & 0 & 0 \\ d_{125} & d_{115} & d_{135} & -d_{145} & 0 & 0 \\ 0 & 0 & 0 & 0 & d_{145} & \frac{1}{2}(d_{115} - d_{125}) \\ 0 & 0 & 0 & 0 & \frac{1}{2}(d_{115} - d_{125}) & \frac{1}{2}(d_{116} - d_{126}) \end{bmatrix}$$

$$d_{3ij} = \begin{bmatrix} d_{311} & d_{312} & d_{313} & d_{314} & 0 & 0 \\ d_{312} & d_{311} & d_{313} & -d_{314} & 0 & 0 \\ d_{313} & d_{313} & d_{333} & 0 & 0 & 0 \\ d_{314} & -d_{314} & 0 & d_{344} & 0 & 0 \\ 0 & 0 & 0 & 0 & d_{344} & d_{314} \\ 0 & 0 & 0 & 0 & d_{314} & \frac{1}{2}(d_{311} - d_{312}) \end{bmatrix}.$$

**Coupling between a Traveling Acoustic Wave and a Bias Electric Field**

An applied electric field affects the traveling acoustic wave by tuning the elasticity of the material. When the applied electric field is small, such tuning in elasticity can be treated by the perturbation theory. The wave equation for a guided acoustic mode is

$$-\rho \omega^2 \, \mathbf{u} = \nabla \cdot (\mathbf{C}\, \boldsymbol{\epsilon}),$$

where $\rho$ is the mass density of the material, $\mathbf{u}$ is the displacement field, $\omega$ is the angular frequency of the eigenmode at given wavenumber $k$.

For a guided acoustic mode, the first-order shift in the eigenfrequency at the given wavenumber $k$ due to the perturbation of elasticity $\Delta \mathbf{C}$ is given by

$$\frac{\Delta \omega}{\omega} = \frac{\int dr \, \Delta c_{ij} \epsilon_i^* \epsilon_j}{2 \int dr \, c_{ij} \epsilon_i^* \epsilon_j} = \frac{\int dr \, d_{kij} E_k \, \epsilon_i^* \epsilon_j}{2 \int dr \, c_{ij} \epsilon_i^* \epsilon_j}.$$

The integral is over the whole cross section that perpendicular to the acoustic propagation direction. Further, the change of wavenumber $\Delta k$ at certain mode frequency $\omega$ is calculated by the dispersion relation of the acoustic mode,

$$\frac{\Delta k}{k} = \frac{\Delta \omega}{\omega} \frac{v_p}{v_g},$$

where $v_p$ and $v_g$ are the phase and group velocity of the guided acoustic mode. The overall acoustic phase change over length $L$ due to the applied electric field is $\Delta k\, L$.

**Acknowledgments:** We thank Prof. Cheng Wang and Cleaven Chia for fruitful discussion. This work is supported by ONR QOMAND grant no. N00014-15-1-2761, DOE HEADS-QON grant no. DE-SC0020376, NSF grant no. DMR-2004536, the Welch Foundation Grant F-1814, and NSF RAISE/TAQS grant no. NSF ECCS-1839197. N.S. is supported by the Natural Sciences and Engineering Research Council of Canada (NSERC), the AQT Intelligent Quantum Networks and Technologies (INQNET) research programme and the DOE/HEP QuantISED programme grant and DOE award no. DE-SC0019219. D. Z. is supported by the Harvard Quantum Initiative (HQI) postdoctoral fellowship and A*STAR SERC Central Research Fund (CRF).

**Author contributions: Linbo Shao:** Conceptualization, Methodology, Investigation, Formal analysis, Visualization, Writing - Original Draft. **Di Zhu**: Methodology, Investigation, Writing - Original Draft. **Marco Colangelo**: Investigation, Writing - Review & Editing. **Daehun Lee**: Investigation, Writing - Review & Editing. **Neil Sinclair**: Methodology, Investigation, Writing - Original Draft. **Yaowen Hu**: Writing - Review & Editing. **Peter T. Rakich**: Writing - Review & Editing. **Keji Lai**: Resources, Methodology, Writing - Review & Editing, Supervision. **Karl K. Berggren**: Resources, Writing - Review & Editing, Supervision. **Marko Loncar**: Resources, Writing - Review & Editing, Supervision.

**Competing interests:** M.L. is involved in developing lithium niobate technologies at HyperLight Corporation. The other authors declare no competing interests.

**Data and materials availability:** All data supporting the conclusion is available in the main text or the Supplementary Information.

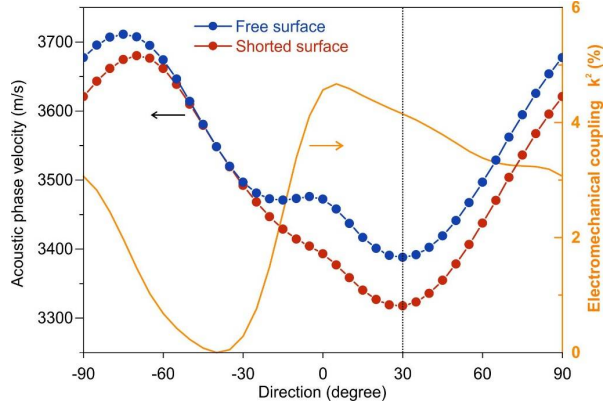

**Extended Data Fig. 1 | Simulated acoustic phase velocities for varying directions on X-cut LN**. The direction is defined by the angle respective to the crystal Z axis. The electromechanical coupling coefficient $k^2 = 2\,(v_o - v_m)/v_o$, where $v_o$ and $v_m$ are the phase velocities when the top surface is free and electrically shorted, respectively. The direction of the waveguide used in our device is 30°, as indicated by the dash line.

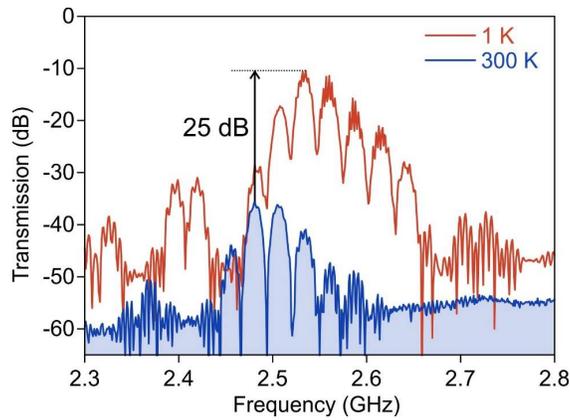

**Extended Dara Fig. 2 | Measured transmission spectra of the acoustic modulator at temperatures of 300 and 1.3 K.** The results indicate a 25 dB improvement in peak transmission at low temperature. The frequency shift of the spectrum is due to the temperature dependent elasticity of LN.

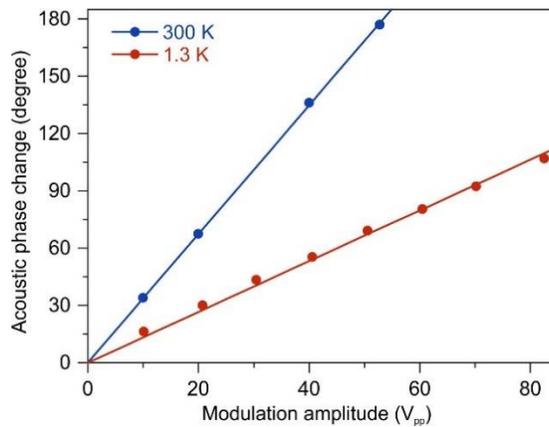

**Extended Data Fig. 3 | Measured Peak-to-peak acoustic phase changes due to sinusoidal modulating signals of varying peak-to-peak voltage ($V_{pp}$) at room and cryogenic temperatures.** The sinusoidal modulating signals are of the frequency $f_{mod}$ =10 kHz. Linear fits show $V_\pi$ of 53 V at room temperature (300 K) and 135 V at 1.3 K, respectively. The same device is measured as that in Fig. 2.

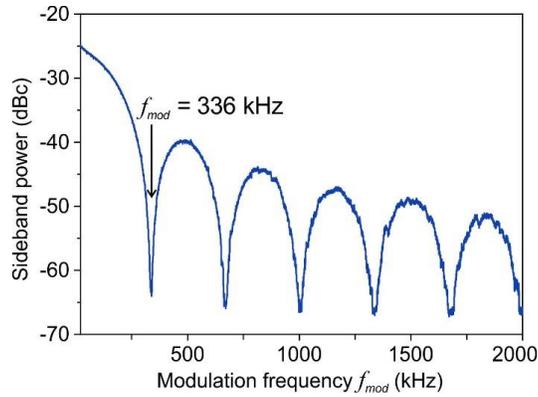

**Extended Data Fig. 4 | Modulation bandwidth of the 1-cm-long electro-acoustic phase modulator.** The modulation efficiency is indicated by the first sideband power due to the phase modulation. The measured 3-dB bandwidth is 110 kHz. The modulation approaches zero at $f_{\text{mod}}$ = 336 kHz when the acoustic traveling time through the modulator equals $1/f_{\text{mod}}$. The same device is measured as that in Fig. 2.

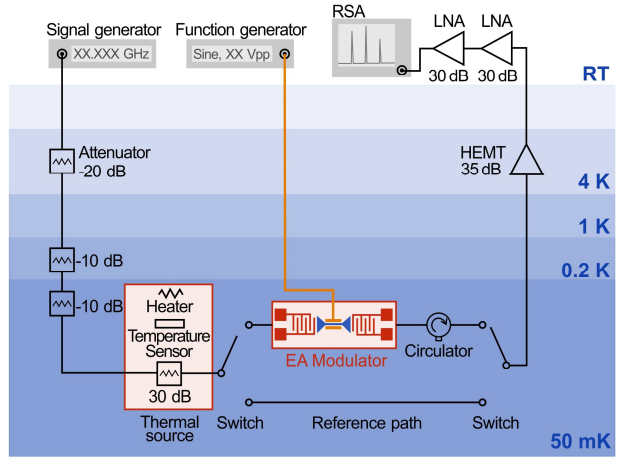

**Extended Data Fig. 5 | Fifty millikelvin measurement setup.** The electro-acoustic modulator is mounted on a mixing plate of a dilution fridge with a base temperature below 50 mK. The input microwave signal is provided by a signal generator and, to ensure negligible contribution of thermal noise, passed through attenuators at various temperature stages of the fridge before going into our device. The output microwave signal from the modulator passes through a circulator, a high-electron-mobility transistor (HEMT) amplifier at 4 K, two low-noise amplifiers at room temperature, and is finally detected by a real-time spectrum analyzer (RSA). The modulation signal is provided by a function generator. A controlled and thermally isolated thermal source, which consists of a heater, temperature sensor, and a 30 dB attenuator, is installed in the microwave line before our electro-acoustic modulator to calibrate the gain and added noise in the output/readout line.

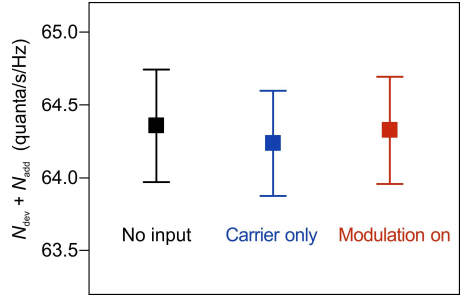

**Extended Data Fig. 6 | Noise measurement of the electro-acoustic modulator at 50 mK.** Total noise power spectrum density near $f_c$ when (1) no signal is applied to the electro-acoustic modulator (black), (2) only the carrier microwave signal is applied (blue), and (3) both the carrier microwave and modulation signals are applied (red). The total noise power $N_{tot} = N_{dev} + N_{add}$, where $N_{dev}$ is the noise of the electro-acoustic modulator and $N_{add}$ is the added noise from the readout chain. $N_{add}$ is mainly determined by the high-electron-mobility transistor (HEMT) at the 4 K stage. The electro-acoustic modulation adds negligible noise and is thus suitable for quantum phononics. The same device is measured as that in Fig. 2.

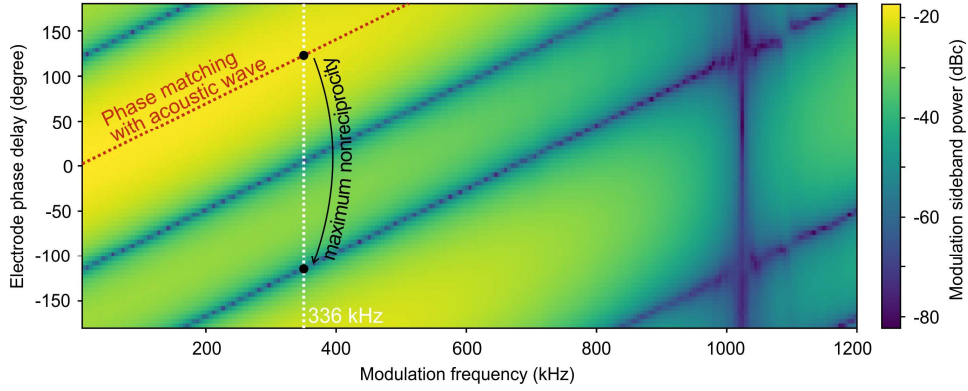

**Extended Data Fig. 7 | Phase matching between a traveling acoustic wave and a quasi-traveling electric field.** The modulation sideband power is measured with varied modulation frequency and phase delay between the three electrodes. The maximum modulations that satisfy the phase matching condition are indicated by the red line. The condition for the maximum nonreciprocity in phase modulation is indicated by the black dots, as the counter propagating acoustic waves experience opposite phase delays compared to the propagating waves. The sideband power is normalized to the unmodulated carrier acoustic wave power. The measured date is from the same device as in Figs. 4C and 4D, which consists of three electrodes with overall modulation length of 1 cm.